# Can Population III Stars at High Redshifts produce GRB's?

C. Sivaram and Kenath Arun
Indian Institute of Astrophysics, Bangalore

**Abstract:** Gamma-ray bursts are the most luminous physical phenomena in the universe, consisting of flashes of gamma rays that last from seconds to hours. There have been attempts to observe gamma ray bursts, for example, from population III stars of ~500 $M_{\odot}$ at high redshifts. Here we argue that collapse of such high mass stars does not lead to gamma ray burst as their core collapse temperatures are not sufficient to produce gamma rays, leading to GRBs.

It is now believed that the long-duration gamma ray bursts (> 2 sec) are associated with the beamed energy from a specific kind of hypernova event, such as the death of supermassive stars once its silicon "burning" is complete with a Zero Age Main Sequence (ZAMS) mass between 40 and 100 solar masses causing a direct collapse of the core to a black hole [1, 2]. The close connection between GRB's and Type Ib/c supernovae shows that the progenitor stars are almost exclusively low metallicity (at ZAMS) fast rotating Wolf-Rayet stars. [3, 4]

The thermal energy associated with a black body at temperature T is given by:

$$E = \frac{4}{3}\pi a T^4 R^3 \quad \text{... (1)}$$

Where, R is the size of the compact region (close to that for a black hole)

If the size is compact enough at a certain temperature, the gravitational energy, given by $\frac{GM^2}{R}$, associated with the dominant black body radiation will also be significant. The total energy, mass energy plus gravitational, is then given by: [5]

$$E = \frac{4}{3}\pi a T^4 R^3 - G\left(\frac{4}{3}\pi a T^4 \Big/ c^2\right)^2 R^5 \quad \text{... (2)}$$

(Putting $M = \frac{4}{3}\pi a T^4 R^3$)

As the gamma rays observed correspond to a temperature of T~$10^{12}$ K, for the compact opaque region, we estimate the size R for which the two quantities in equation (2) become comparable.

This gives:

$$R = \left( \frac{3}{4\pi} \frac{c^4}{aT^4 G} \right)^{1/2} \approx 200\,km \text{ ... (3)}$$

as the radius of the compact region, which following the usual scenario for gamma ray bursts, would correspond to be about the Schwarzschild or gravitational radius. Equation (3) corresponds to a black hole mass of the order of 60-70 $M_{\odot}$ for a required temperature of $\sim 10^{12}$ K to produce gamma rays.

This puts an upper limit on the size of stars that can collapse to give a GRB as $R_S \propto M$. The upper limit (for T$\sim 10^{12}$ K) of about 60 solar masses corresponds quite well to the mass of the supposed progenitor stars of GRBs, i.e. Wolf-Rayet stars, in the usual scenario.

The particles in the system will be in thermal equilibrium with the black body radiation. Therefore, similar analysis with gas thermal energy $(Nk_B T)$ yields a similar result.

Equation (3) implies the significant result that much larger stellar masses corresponding to the primeval population III stars [6, 7] (i.e. 500 to $10^3$ $M_{\odot}$ (or greater)), would result in a much smaller temperature, prior to the collapse into a black hole. This would in turn imply that gamma rays could not be produced by collapse of such stars. So future searches for gamma ray bursts (at higher z) would be perhaps constrained as to the type of progenitors which could produce such events.

The argument can also be extended to the case of neutrino pair production at high temperatures. Like in a typical core collapse SN (like SN1987A) most of the gravitational binding energy is expected to be released in the form of neutrinos (and antineutrinos) of all three flavours. [8]

The core temperature of the 'proto-neutron star' is around $10^{12}$ K (it may collapse to a black hole subsequently!). For the case of neutrinos, the black body formula for energy density would be: [9]

$$\varepsilon_{rad} \approx \frac{7}{16} aT^4 \text{ ... (4)}$$

For three flavours, it would be:

$$\varepsilon_{rad} \approx \frac{7}{16} aT^4 \times 3 \times 2 \text{ ... (5)}$$

This when used in equations (2) and (3) would lead to the radius R as:

$$R = \left( \frac{3}{4\pi} \frac{c^4}{\left( \frac{7}{16} aT^4 \times 3 \times 2 \right) G} \right)^{1/2} \approx 125km \text{ ... (6)}$$

So this implies for T~$10^{12}$ K, a mass of ~40 $M_{\odot}$.

This again would constrain the type of progenitor which could produce simultaneously both gamma rays and neutrinos. Perhaps only short duration bursts, caused by merger of two neutron stars or tidal break up of a NS can lead to such simultaneous intense bursts of both gamma rays and neutrinos. [10]

In any case massive population III stars above 250 – 300 $M_{\odot}$ are expected to collapse into black holes [8] and those in the range 140 – 250 $M_{\odot}$ are expected to undergo the so called Pair Instability Super Nova, i.e. PISN, where production of $e^+e^-$ pairs at around the oxygen burning temperature at $2 \times 10^9 K$ would reduce the radiation pressure by $\frac{1}{8} aT^4$ and cause a collapse followed by complete fragmentation of the star. Above 250 $M_{\odot}$, the collapse terminate in a black hole, which by the above argument would not produce a gamma ray burst (or a neutron star).

**Reference:**

1. T. Piran, *Phys. Rep.*, 314, 575, 1999
2. G. Chincarini *et al*, *The Messenger*, 123, 54, 2006; A. MacFadyen and S. Woosley, *Ap J.*, 524, 262, 1999
3. R. G. Detmers *et al*, arXiv:0804.0014v1 [astro-ph], 31 Mar 2008
4. A. Marle *et al*, *A&A*, 469, 941, 2007
5. C. Sivaram, *Ap&SS*, 140, 403, 1988
6. C. Sivaram, *Ap&SS*, 86, 501, 1982; *Ap&SS*, 88, 507, 1982

7. C. Sivaram, In Relativistic Astrophysics and Cosmology, Edited by V. de Sabbata and T. Karade (World Scientific), p.228, 1984
8. H. Kreckel *et al*, *Science*, 329, 69, 2010; A. Heger and S. Woosley, *Ap J.*, 567, 532, 2002
9. A. Gal-Yam *et al*, *Nature*, 462, 624, 2009
10. V. Bromm, *Science*, 329, 45, 2010
11. C. Sivaram and Kenath Arun, arXiv:0911.2747v1 [astro-ph.CO], 14 Nov 2009